 \documentclass[preprint,review,12pt]{elsarticle}

\usepackage{upgreek}
\usepackage{amsmath,amssymb} 
\usepackage{enumerate}

\usepackage{graphicx}

\usepackage[usenames, dvipsnames]{color}

\journal{Ultramicroscopy}
\begin{document}

\begin{frontmatter}



\title{Electrostatic electron mirror in SEM for simultaneous imaging of top and bottom surfaces of a sample}

\author[label1]{Navid Abedzadeh}
\author[label2]{M. A. R. Krielaart}
\author[label1]{Chung-Soo Kim}
\author[label1]{John Simonaitis}
\author[label3]{Richard Hobbs}
\author[label2]{Pieter Kruit}
\author[label1]{Karl K. Berggren}

\address[label1]{Research Laboratory of Electronics, Massachusetts Institute of Technology, Cambridge, MA 02139, USA}
\address[label2]{Department of Imaging Physics, Delft University of Technology, Lorentzweg 1, 2628CJ Delft, The Netherlands}                                                                                                                                          
\address[label3]{School of Chemistry, Advanced Materials and Bioengineering Research (AMBER) Centre and Centre for Research in Adaptive Nanostructures and Nanodevices (CRANN), Trinity College Dublin, Dublin, Ireland}

\begin{abstract}
The use of electron mirrors in aberration correction and surface-sensitive microscopy techniques such as low-energy electron microscopy has been established. However, in this work, by implementing an easy to construct, fully electrostatic electron mirror system under a sample in a conventional scanning electron microscope (SEM), we present a new imaging scheme which allows us to form scanned images of the top and bottom surfaces of the sample simultaneously. We believe that this imaging scheme could be of great value to the field of in-situ SEM which has been limited to observation of dynamic changes such as crack propagation and other surface phenomena on one side of samples at a time. We analyze the image properties when using a flat versus a concave electron mirror system and discuss two different regimes of operation. In addition to in-situ SEM, we foresee that our imaging scheme could open up avenues toward spherical aberration correction by the use of electron mirrors in SEMs without the need for complex beam separators.

\end{abstract}


\end{frontmatter}


\section{Introduction}

In its simplest form, an electron mirror creates an electric field in which incident electrons slow down to a complete standstill before being accelerated away in the opposite direction. Multi-electrode electron mirrors have been used in mirror electron microscopy (MEM) \cite{Hottenroth1936,Hottenroth1937,Orthuber1948,Oman1969ElectronMicroscopy,Lukyanov1974} and low-energy electron microscopy (LEEM) \cite{Telieps1985LEERM,HayesGriffith1991HistoricalTechn, TrompLEEM1991, Tromp2000LEEM}. Today, aside from LEEM, the most prominent application of electron mirrors is in chromatic and spherical aberration correction \cite{Rempfer1990, Rempfer1997, Tromp2010}. More recently, designs for novel electron microscopy schemes that lower beam-induced damage to fragile biological samples have been proposed. Under the banner of quantum electron microscopy (QEM), two such designs have been proposed that are based on electron cavities consisting of two electron mirrors between which electrons would reflect back and forth to probe a sample multiple times \cite{Kruit2016DesignsMicroscope,Juffmann2017,Turchetti2019}. Implementation of electron mirrors in SEMs has not been widely reported. In this work, we explore the use of tunable, easy to construct electron mirror systems in an SEM under different imaging regimes.

\begin{figure}
	\centering
	\includegraphics[width=1\textwidth]{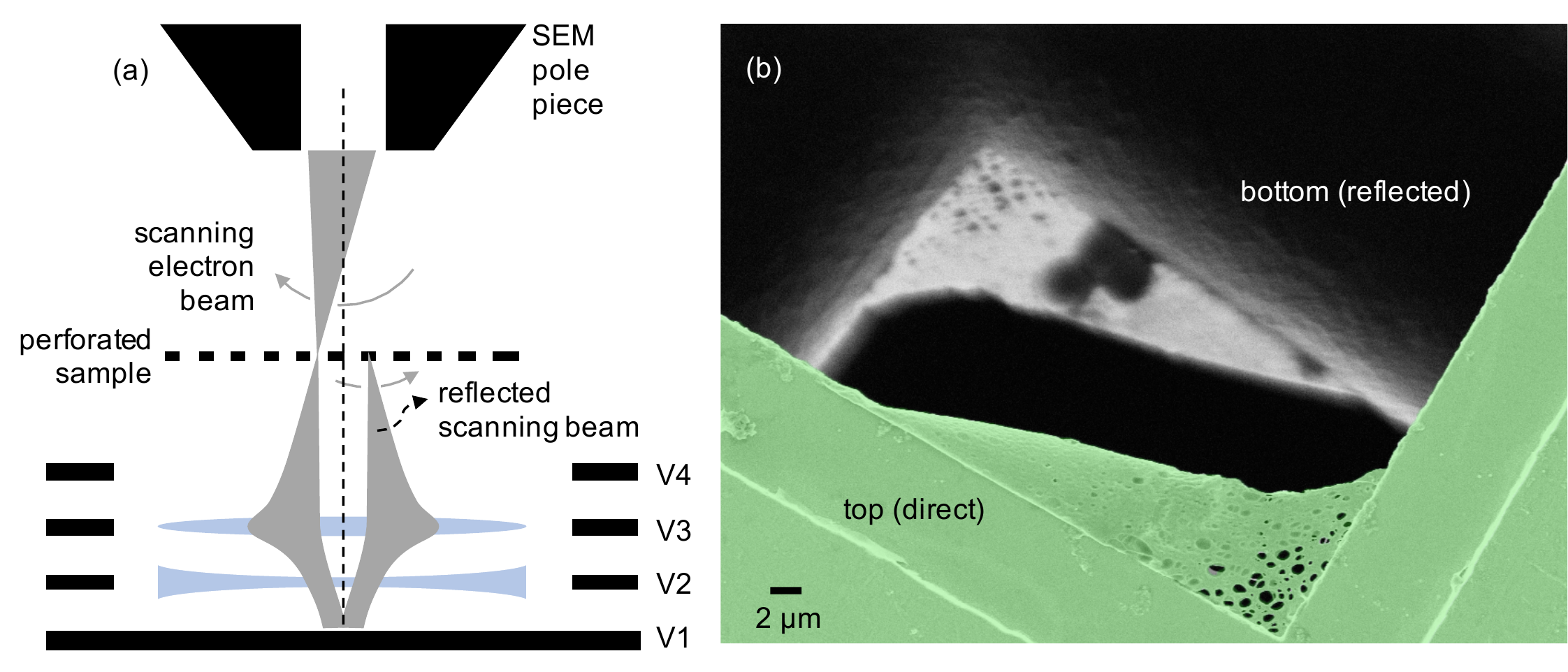}
	\caption{Simultaneous imaging of the top and bottom surfaces of a sample in SEM. (a) Schematic of the optical design and electron ray trajectory: the tetrode mirror system, comprised of electrodes V1-V4, reflects a scanning electron beam where it is not obstructed by the perforated sample, and refocuses it on the bottom surface of the sample. The schematic is not drawn to scale and the light optics lenses (blue) are drawn only as an analogy. (b) SEM micrograph showing the top (colorized in green) and bottom (greyscale) surfaces of a sample within the same frame. The sample shown here consists of a suspended holey carbon membrane that is attached to a TEM copper grid. The reflected image shows particles on the bottom side of the membrane which are not visible in the direct image.}
	\label{Fig1}
\end{figure}

During the development of electron mirrors for the QEM project, we incorporated a fully electrostatic multi-electrode mirror in a conventional SEM, and found a new imaging scheme in which it is possible to produce simultaneous scanned images of top and bottom surfaces of perforated samples. As shown in Fig. \ref{Fig1}(a) a focused electron beam scans a perforated sample, e.g. a transmission electron microscope (TEM) copper mesh. Where the sample obstructs this scanning probe, a scanned image of the top surface of the sample is produced. At scan positions where the probe is not obstructed, the electrons pass through the sample plane towards the electron mirror system placed below. This electron mirror system which consists of a mirror electrode and three electrodes to form a lens is commonly referred to as a tetrode mirror and in this case reflects and refocuses the beam back on the sample plane. Where this reflected scanning probe strikes the bottom surface of the sample, a scanned image of the bottom surface of the sample is produced in the same micrograph. This is due to the fact that the greyscale value of each pixel in an SEM micrograph is dependent only on the number of secondary electrons detected while the beam was dwelling over that pixel as opposed to the source of those secondary electrons. The end result is an SEM micrograph in which the top and bottom surfaces of the sample are simultaneously visible as shown in Fig. \ref{Fig1}(b). It must be noted that we employ the term ``simultaneous" loosely here; depending on the scan speed, there is microseconds to milliseconds between when the top and the bottom images form. In this work, we refer to the image of the top surface as the direct image and the image of the bottom surface as the reflected image. 

A similar imaging technique in a different context was demonstrated by Crewe et al. using custom made scan coils and a magnetic lens \cite{Crewe2000}. In our approach, we take advantage of a fully electrostatic tetrode mirror, which could be easily retrofitted in most conventional SEMs. Furthermore, since the operation of our mirror does not rely on a magnetic immersion lens, the sample could be held at a nominally field-free region which is of interest for investigation of magnetic samples.

The most immediate application of our imaging technique is its potential use in in-situ SEM where it is of interest to image dynamic changes such as crack propagation or melting of a sample \cite{Andersson2004,Torres2011}. Currently, in-situ SEM is limited to capturing changes on one side of the sample at a time. The multi-electrode mirror and the imaging scheme described in this work could be used to simultaneously capture dynamic changes on both surfaces of a thin sample. Incorporation of this mirror system in an SEM chamber could be easily achieved due to its relatively small form factor and ease of assembly and operation. For achieving the highest possible resolution, careful design, machining and assembly of multi-electrode electron mirrors are crucial. We demonstrate that even without meeting those stringent requirements, a simple electron mirror system constructed with off-the-shelf components is nonetheless capable of reflecting the beam and forming a scanning focused probe of about 100 nm diameter to produce reflected images.

Another potential application for our imaging scheme is mirror aberration correction. Conventional mirror aberration correction has relied on separation of the incident (uncorrected) and the reflected (corrected) beams by means of either Y-separators \cite{Rempfer1990,Rempfer1997} or 90-degree separators \cite{Wan2006}. These separators often introduce chromatic dispersions of their own to the beam. Although later designs, notably one by R. Tromp \cite{Tromp2010} eliminated these chromatic dispersions, such beam separators remain complex and are generally not designed for use in SEMs. More recently, Dohi and Kruit showed a promising design in which two miniaturized electron mirrors along with small-angle deflectors are used in order to correct the aberrations of an SEM objective lens \cite{Dohi2018}. However, their approach requires significant modifications to the electron column.

Our imaging scheme could lead to a fundamentally different approach to mirror aberration correction by eliminating the need for deflectors of any kind. In a similar imaging setup as shown in Fig. \ref{Fig1}(a), one could replace the flat mirror electrode (V1) with an aperture electrode in order to create a tunable curved mirror potential. With the mirror voltages tuned correctly, one could image the bottom surface of a perforated sample with a spherical- and/or chromatic-aberration-corrected probe without the need for beam separators. This simple approach comes with a number of limitations chief among which is the need for samples that do not block the entire field of view. It must be noted that although we demonstrate image resolution improvements when using a concave five-electrode mirror system, we do not attempt rigorous aberration correction which is an involved process and hence beyond the scope of this work.

\section{Experimental design}

To produce simultaneous scanned images of the top and bottom surfaces of a sample in an SEM, maintaining focus at the sample plane for the beam on its way down as well as its way up is a crucial requirement. To achieve this requirement, two electron optical elements in addition to the ones the SEM is already equipped with are needed: (1) an electron mirror, and (2) an electron lens that is positioned between the mirror and the sample plane. The mirror reflects the incident beam and redirects it towards the sample plane while the lens refocuses the reflected beam. Depending on the strength of the lens element, however, this requirement could be met in one of two different regimes of operation. These regimes are defined by whether the back-focal plane or the image plane of the mirror system coincides with the sample plane. Furthermore, the mirror element could be flat or curved depending on the intended application and voltage restrictions.

In this work, as shown in Fig. \ref{Fig1}(a), we refer to the mirror system electrodes, as V1 through V4 from bottom to top. Electrode V1 is a flat stainless steel plate to which a negative bias larger in magnitude than that of the acceleration voltage of the beam is applied. Electrode V2, sometimes referred to as the cap electrode, is an electrode with an aperture at its center. It acts as a field limiting aperture for the mirror electric field and due to the curved field lines near its aperture, effectively behaves as a diverging lens \cite{Lukyanov1974, Rempfer1989}. For simplicity, we keep this electrode grounded. The role of electrode V3 is to provide a positive lensing effect, allowing for refocusing of the reflected electron beam onto the bottom surface of the sample. Hence, we refer to this electrode as the lens electrode. Finally, the grounded electrode V4 completes the lens component of the tetrode mirror and minimizes the amount of stray fields from the electrodes below.

\subsection{Mechanical considerations}
\begin{figure}[h] 
	\centering
	\includegraphics[width=0.6\textwidth]{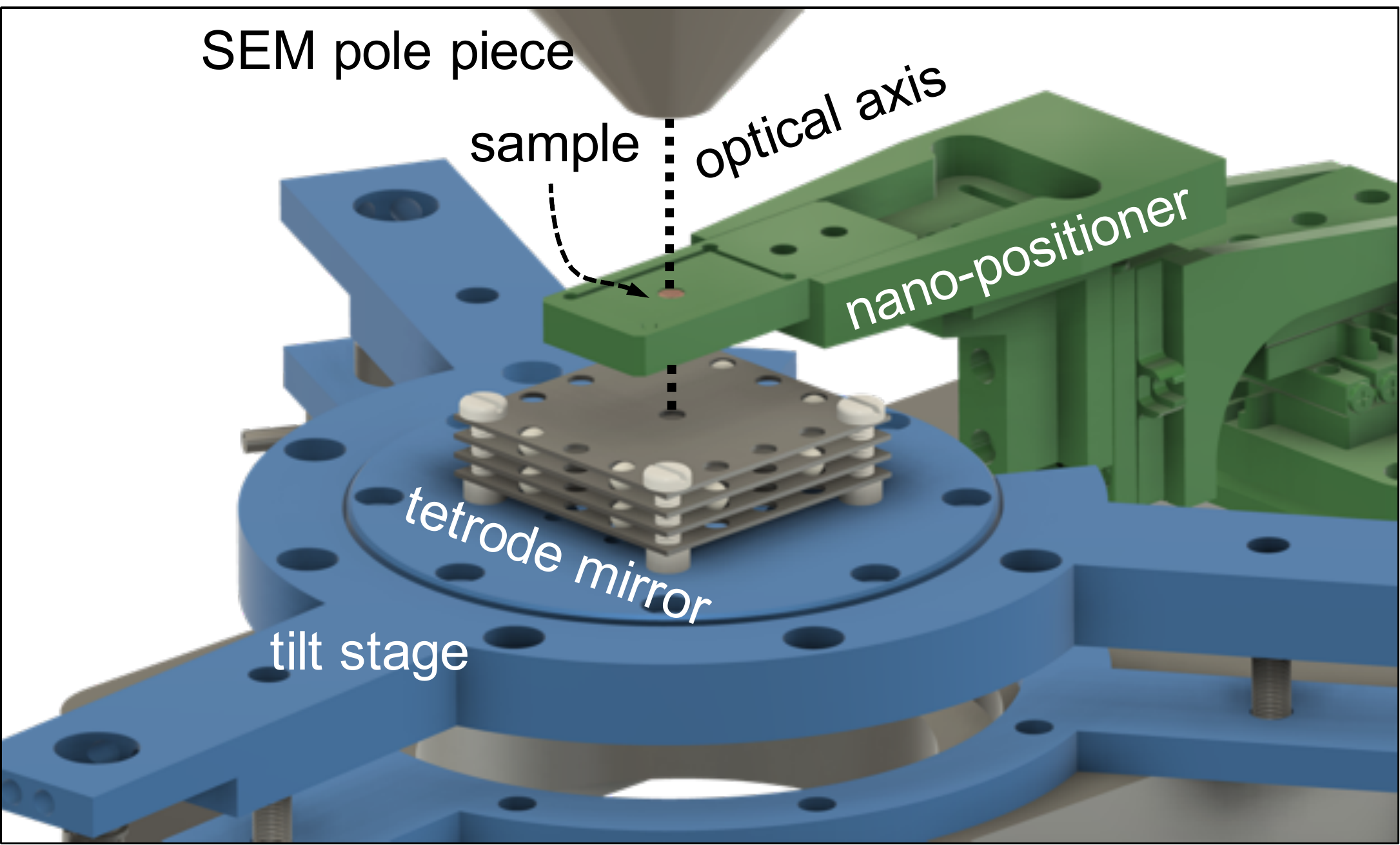}
	\caption{Experimental setup: 3D CAD diagram of a perforated sample (copper grid) mounted on a three-axis nano-positioner (green). The sample is held above a tetrode electron mirror (gray) which is mounted on a tilt stage (blue). Both the nano-stage and the tilt stage are mounted on a three axis translation stage which is connected to the SEM door (not shown). The optical axis of the tetrode mirror is aligned with that of the SEM objective lens. The square plates (gray) used in the mirror assembly are 5 cm on each side.}
	\label{Fig2}
\end{figure}

The form factor of the electron mirror system is dictated by the available space inside the SEM, mechanical and machining constraints, as well as electric field and electron optical requirements. With these considerations, we opted to use commercially available stainless steel plates with regularly spaced through-holes in order to construct the mirror stack. Figure \ref{Fig2} shows these plates in grey, assembled using plastic bolts and alumina balls. The plates, acquired from Kimball Physics Inc., are  0.635 mm (0.025 inches) thick and the through-holes are 3.175 mm (0.125 inches) in diameter, including the central hole which is the aperture through which the electron beam passes. The alumina balls provide electrical insulation between the electrodes, as well as stability to the structure by fixing the plates in the lateral direction under the vertical pressure from the bolts. We used alumina balls of radius 3.5 mm which set the separation between the electrodes to about 1.35 mm. The minimum distance between the sample plane and the mirror electrode is set by the thickness of the mirror stack and the sample holder and in our case is about 25 mm. The distance between the sample and the SEM objective lens (working distance) was set to about 7 mm.

The alignment of the mirror stack with respect to the SEM objective lens is crucial to the performance of the electron mirror. For this alignment, we made use of separate translation and tilt stages, as shown in Fig. \ref{Fig2}. The sample was mounted on a dedicated 3-axis nano-positioner which provided independent positioning of the sample relative to the mirror stack. In order to ensure that the mirror is level, the mirror stack was mounted on a piezo-controlled tilt stage. Both the nano-positioner and the tilt stage were in turn mounted on a 3-axis translation stage which is attached to the SEM door inside the vacuum chamber. 

We used a variety of samples in our experiments. However, in all cases these samples were only partially obstructing the field of view, allowing for electrons to pass through undisturbed at some scan positions. Furthermore, all samples were relatively thin (tens of micrometers) and had nanometer to micrometer features to resolve during imaging. Among these samples were TEM copper grids of varying mesh sizes, and holey carbon films on copper grids.

The SEM used in our experiment is a Zeiss LEO 1525 with a Schottky electron source with an energy spread of about 1 eV. To allow for construction of taller experimental setups inside the SEM chamber, we replaced the original SEM door and the 5-axis stage with a custom door and a 3-axis translation stage with a smaller form factor. The appropriate flanges for high-voltage connection in order to bias the mirror electrodes were incorporated in the custom SEM door.

\subsection{Electron optical considerations}
Electric field and electron ray trajectory simulations performed on the commercial software Lorentz 2E provided insight into the approximate voltages for each electrode in the mirror system. From there, minor voltage adjustments were needed in order to obtain an in-focus micrograph. For consistency, we performed all experiments in this work at the beam acceleration voltage of 3 kV, beam current of about 10 pA, and beam convergence angle of about 2 mrad. The upper limit for the beam acceleration voltage is set by how large of a voltage could be safely applied to the mirror system.

The voltage applied to the mirror electrode defines the plane of reflection. In the case of the flat tetrode mirror, a schematic of which is shown in Figs. \ref{Fig1}(a) and \ref{Fig2}, the magnitude of this voltage must be at least as large as the beam acceleration voltage to avoid collision between the electrons and the mirror electrode. In practice, we must account for the inherent roughness in the surface of the stainless steel mirror plate which leads to a non-uniform electric field near the surface. We set the magnitude of this negative bias to 20-50 V above the acceleration voltage of the electron beam in order to ensure that the beam is reflected on a flat potential surface. Failure to do so could result in distortion and loss of image resolution. In the case of the five-electrode aperture mirror used for curved mirror operation, a much larger negative voltage applied to the mirror electrode is required in order to achieve a potential more negative than beam acceleration voltage at the center of the aperture.

The cap electrode V2 was grounded in order to terminate the electric field lines formed above the mirror electrode V1. In certain applications, it is beneficial to apply a positive bias to the cap electrode in order to increase the magnitude of the electric field in the mirror region. However, in our experiment, high electric fields are not crucial and we opted for a grounded cap electrode which lowers the risk of electrical breakdown between the mirror and the cap electrodes. 

The voltage applied to the lens electrode (V3) controls the convergence angle of the beam entering and exiting the mirror system. It is worth noting that the effects of the lens voltage are not fully decoupled from the cap electrode (V2). In other words, an increase in the positive voltage of the cap electrode leads to more severe diverging action which must be counteracted with a larger voltage on the lens electrode to maintain a fixed focal length. By applying a negative voltage to the lens electrode (V3) we opted for a decelerating lens which achieves the same focal length at a smaller voltage compared to an accelerating lens, reducing the risk of electric breakdown in the setup. 


Typical factors limiting SEM image resolution include the probe size, sample material, acceleration voltage and for slow scans, beam and stage stability \cite{Carter2016}. In our imaging scheme, since the sample and the beam energy for the incident and the reflected beams are the same, the image resolution comes down to the probe size before and after reflection. With the added complexity of the electron mirror system as well as the added optical path in our imaging scheme, the reflected beam is expected to suffer from various additional aberrations and distortions beyond the ones that the SEM objective lens imparts on the incident beam. Therefore, it would be reasonable to expect a larger probe size and hence poorer resolution of the reflected image (bottom) compared to the direct (top) image. The aberrations of the tetrode mirror can be kept relatively small when the spacing between the mirror and cap electrodes is small compared to the radius of the aperture in the cap electrode \cite{Krielaart2021}; however, this requires the use of micromachined optics which we do not pursue in this work, as we focus here on an ``off-the-shelf" approach.

\section{Results and discussion}
In this section, we discuss imaging under different electron optical regimes defined by the trajectory of the reflected beam as determined by the strength of the lens element within the mirror system. Furthermore, we present reflected images produced with a curved mirror constructed with five electrodes. We analyze image properties associated with each regime and experimental setup. 

\subsection{Broad beam over the mirror surface}
\begin{figure}
	\centering
	\includegraphics[width=1\textwidth]{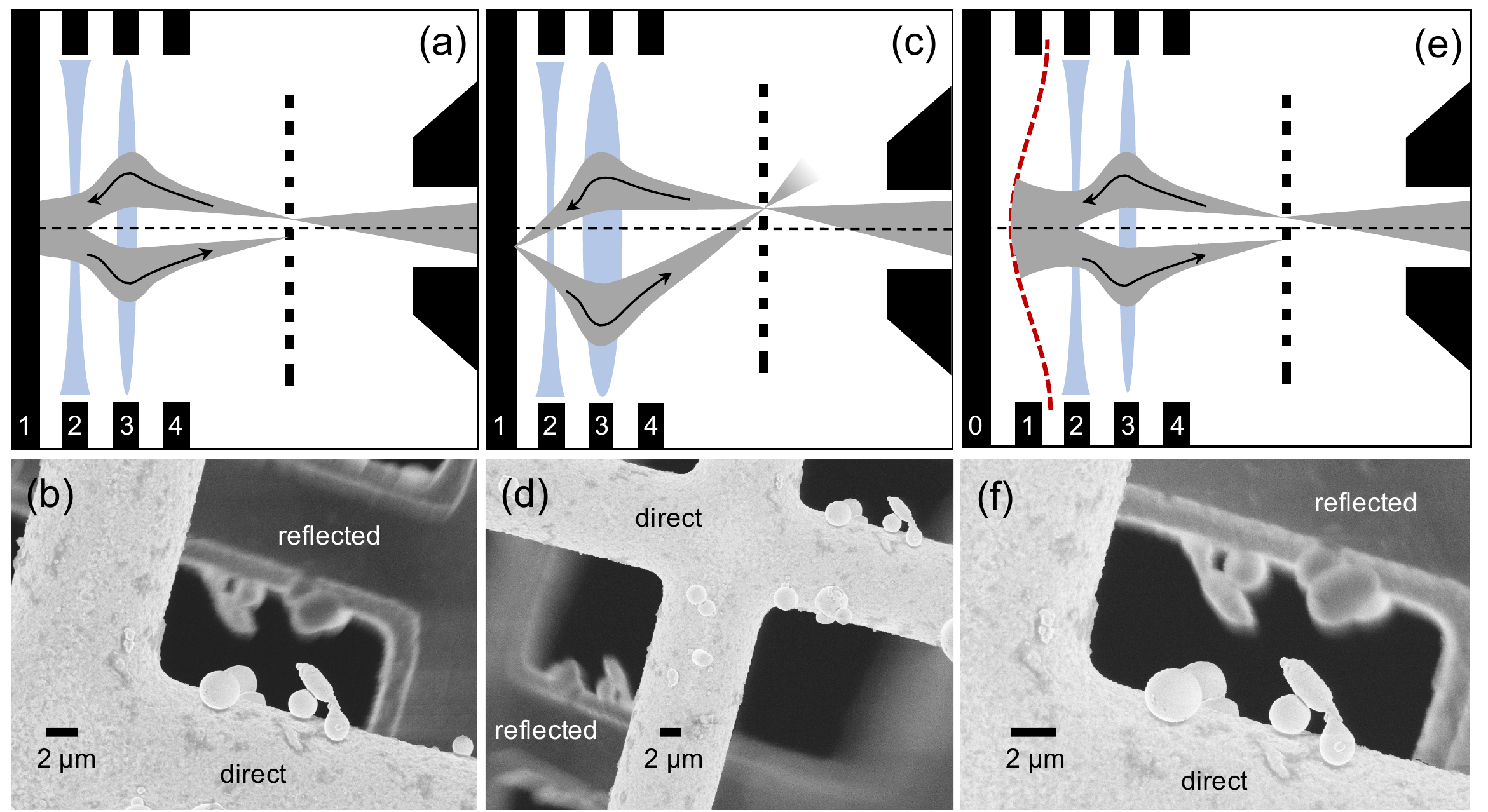}
	\caption{Schematic electron trajectories and their corresponding SEM micrographs obtained under different regimes of mirror imaging. (a)-(b) Tetrode mirror voltages tuned such that the back-focal plane of the mirror system coincides with the sample plane (dashed line) resulting in an SEM micrograph with point symmetry between the direct and reflected images. (c)-(d) Tetrode mirror voltages tuned such that the image plane of the mirror system is conjugated with the sample plane. Note that the micrograph in (d) demonstrates a misaligned experimental setup while the schematic in (c) corresponds to a perfectly aligned system such that the incident and the reflected beams pass through the exact same spot on the sample plane. (e)-(f) Five-electrode mirror with voltages tuned such that the incident beam is reflected on a curved potential surface (red dashed line). Careful engineering of the curved potential could eliminate the spherical aberration of the objective lens in the reflected beam (not attempted here). Schematics are not drawn to scale. All scale bars are 2 $\upmu$m.}
	\label{Fig3}
\end{figure}

The first regime of operation is defined by the tetrode mirror voltages set such that for a given beam energy, the back-focal plane of the mirror and lens coincides with the sample plane. An incoming electron beam that is focused on the sample plane by the objective lens of the SEM will enter the tetrode mirror and consequently be spread out over the mirror electrode. After reflection, the beam will be refocused on the sample plane. Figure \ref{Fig3}(a) shows a schematic of the electron trajectory and Figure \ref{Fig3}(b) shows an example of a micrograph produced under this regime of operation. 

A property of images obtained in this regime is point symmetry about the optical axis between the direct image and the reflected image. The reason is apparent from the electron trajectory schematic: the incident and the reflected beams pass through the sample plane at the same distance to but opposite sides of the optical axis. This is especially useful when performing in-situ microscopy on samples both sides of whom undergo dynamic changes that are of interest.

Ideally, our imaging scheme in this regime would produce the best image resolution over the entire field of view with the SEM settings and the mirror voltages set such that the pivot point of the SEM scan is imaged on the reflection plane. The SEM pivot point is a virtual stationary point below the scan coils around which the scanning beam hinges. By imaging this point at the spot where the optical axis meets the reflection plane, we could ensure that the broad beam remains stationary over the reflection plane near the optical axis as opposed to scanning over the surface of the mirror which could lead to increased aberrations and distortions. Unfortunately, meeting this condition is not trivial in our Zeiss SEM due to software overrides for currents applied to the scan coils and the objective lens.

Since the incident beam is spread out over the mirror plane, this regime of operation is also useful for applications in which manipulations to the electron wave front using an electron mirror is of interest. An example of such application is electron diffraction by means of a diffractive electron mirror \cite{Krielaart2018} which is of interest in novel low-damage electron microscopy schemes \cite{Kruit2016DesignsMicroscope,Okamoto2010}. Another example of an application in which this regime of operation is used is mirror aberration correction which will be discussed in Section \ref{sec:curved}.

\subsection{Focused beam over the mirror surface}
Another mode of operation is realized by increasing the positive lens strength of the mirror system until the image plane coincides with the sample plane. In other words, the lens component images the sample plane onto the mirror electrode, and vice versa. We achieve this regime electrically by keeping all settings constant while increasing the magnitude of the voltage applied to the lens electrode (V3) compared to the voltage required for the first regime discussed above. The electron trajectory schematic for this regime is shown in Fig. \ref{Fig3}(c), from which it becomes clear that the incident and the reflected beams pass through the sample plane at exactly the same point and hence for an opaque sample, no (scanned) image of the bottom surface could form since the object would block the path of the incident beam whose reflection would have otherwise reached the bottom side of the sample. In practice, we only observe simultaneous images of top and bottom surfaces of a sample in this regime when there is a slight misalignment between the optical axes of the mirror system and that of the SEM objective lens. This misalignment could be achieved by a tilt or lateral shift of the mirror. Figure \ref{Fig3}(d) shows an SEM micrograph obtained under this regime when there was a misalignment in the setup. 


A noteworthy difference between the first regime of operation with a broad beam over the mirror and this regime is regarding where the pivot point of the SEM scan is imaged. With the sample plane being imaged over the reflection plane in the latter regime, the pivot point by definition cannot be imaged on the reflection plane, causing the focused beam to scan over the reflection plane. This deviation from the optical axis near the reflection plane leads to significant aberrations and distortions which are apparent in the peripheries of the reflected image in Fig. \ref{Fig3}(d).

Although not very useful for producing scanned images of a sample, this regime of operation is of particular interest for applications such as multi-pass electron microscopy where multiple interactions between a stationary beam and a weak phase object inside an electron cavity could lead to more efficient phase contrast imaging \cite{Juffmann2017}. It is crucial for multi-pass electron microscopy that the reflected beam be re-imaged exactly back on the sample in order for the beam to accrue the correct phase. Although in multi-pass electron microscopy the beam incident on the sample must not be focused, the mirror settings, namely having the sample on the image plane of the tetrode mirror, is identical to what we present in this work.

\subsection{Broad beam over a curved mirror potential} \label{sec:curved}
Appropriately curved electron mirrors have been shown to be capable of inducing spherical and chromatic aberrations with sufficiently negative coefficients in order to cancel out the aberrations of the objective lens \cite{Rempfer1990,Tromp2010}. Here, without attempting to demonstrate quantitative cancellation of spherical aberration, we show that the same imaging technique discussed above could be performed with a curved mirror. To create a curved mirror, we added an aperture electrode to our previous tetrode mirror assembly. In this new five-electrode system, as shown in Fig. \ref{Fig3}(e), a concave reflection plane is formed by applying a negative bias to the aperture electrode V1. Similar to the first regime where the beam is spread over the mirror, the voltages are tuned such that the back-focal plane of the mirror system coincides with the sample plane.

The voltage applied to the flat electrode in this setup (V0) could control the curvature of the reflection potential surface while keeping the mirror voltage (V1) constant. A negative voltage applied to the flat electrode V0 would make the curved mirror potential less concave whereas a positive V0 would create a more concave mirror potential surface. Although, in principle, the flat electrode V0 is not necessary to produce a curved mirror potential, in practice, it proves useful for two reasons: it acts as a flat surface terminating the field lines emanating from the mirror electrode (V1), and also when biased negatively, it allows for an identical mirror potential curvature under a lower V1 voltage for less chance of electric breakdown.

A concave mirror potential has some positive lensing effect which must be accounted for by applying a smaller voltage to the lens electrode (V3) in order to maintain a focused beam on the sample plane. In addition, the curvature of the mirror dictates the amount of spherical and chromatic aberrations imparted on the beam. These considerations create a multi-variable problem and in practice, sweeping the phase space of various voltages in an electron trajectory simulation is the efficient way to ensure minimization of spherical and chromatic aberrations while maintaining a set focal length as shown by Tromp et al. \cite{Tromp2010}.

The reflected image shown in Fig. \ref{Fig3}(f) shows qualitative resolution improvements compared to the reflected image in Fig. \ref{Fig3}(b). We attribute this improvement in image resolution to the lower spherical and chromatic aberrations of the curved mirror system compared to those of the flat mirror system. Although, this is not an exercise in aberration correction, it does point towards our ability to change the aberration coefficients of our mirror system by manipulating the curvature of the reflection surface. It is conceivable that our imaging scheme combined with an accurately tuned curved mirror could be used in the future to correct the aberrations of the SEM objective lens without the need for a beam separator.


\subsection{Reflected image resolution and field of view}
The discrepancy in image resolution and contrast between the direct and the reflected images is observed to various degrees in all regimes of operation as apparent in all micrographs presented in this work. The poorer resolution and contrast of the reflected image is to be expected considering the added spherical and chromatic aberrations that the uncorrected mirror system imparts onto the reflected beam. In this section, we use micrographs to quantify the beam spot size after reflection, comment on the contrast of the reflected image, and discuss the useful field of view and magnification of the reflected image.

\begin{figure}
	\centering
	\includegraphics[width=1\textwidth]{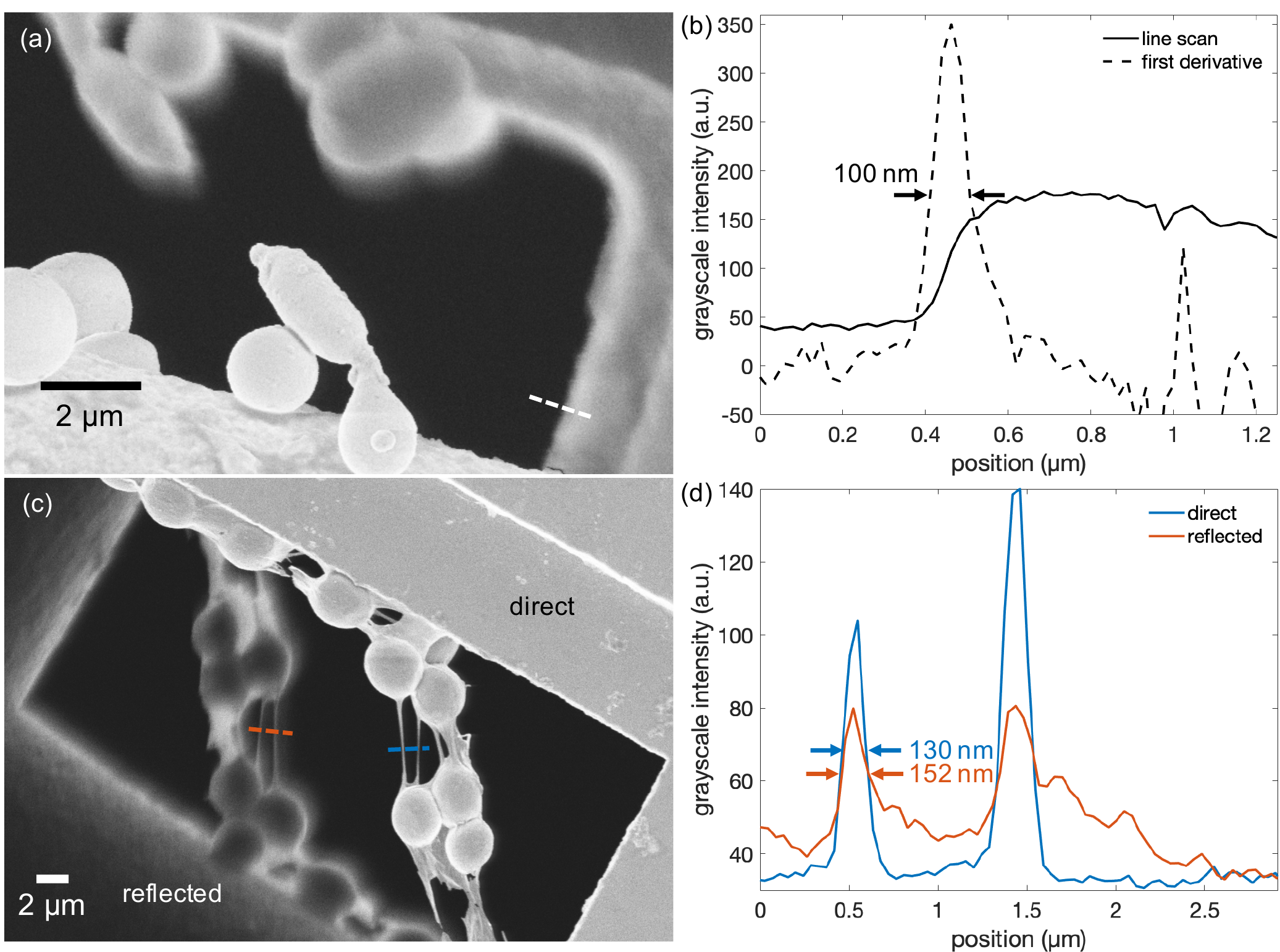}
	\caption{Approximating the reflected probe size and resolution of the reflected image. (a) Reflected image of a sharp edge appropriate for a knife-edge measurement. (b) Image intensity at the reflected image of the sample edge (solid) and its derivative (dashed) measured at the line indicated by the dashed line shown in (a). Note that this is not a measure of the direct beam whose diameter is significantly smaller. (c) Simultaneous imaging of top and bottom surfaces of a sample containing fine features. The dashed lines indicate the lines over which image intensity profiles were obtained for the direct (blue) and the reflected (red) images. (d) The width of one of the carbon strands as measured on the direct image (blue) and reflected image (red) is about 130 nm and 152 nm, respectively.}
	\label{Fig4}
\end{figure}

As a measure of the electron probe size after reflection, we obtained the intensity profile of the edge  indicated by the dashed line in Fig. \ref{Fig4}(a). Figure \ref{Fig4}(b) shows this line profile (solid) along with its first derivative (dashed) after an average filter of window size 3 was applied to the latter for slight smoothening. The full width at half maximum of the derivative indicates a probe diameter of about 100 nm for the reflected beam. This region of the reflected image was chosen for its apparent sharpness and lack of astigmatism compared to the rest of the micrograph. However, factors such as mechanical vibration and the finite sharpness of the edge of the sample make this measurement pessimistic -- the true probe size may be slightly smaller than 100 nm. Note that this measurement is not of the direct beam whose diameter, devoid of the aberrations of the mirror system, is considerably smaller.

To compare the resolution of the direct and the reflected images, we imaged samples with relatively fine features. We deposited micrometer-sized tin particles on a TEM copper grid with a thin carbon membrane on top. In certain areas, the carbon membrane ruptured and wrapped around the tin particles. The result, as shown in Fig. \ref{Fig4}(c), is suspended tin particles wrapped in thin strands of carbon. The blue and red dashed lines represent the line scans whose profiles are shown in Fig. \ref{Fig4}(d). The width of one of the carbon strands as measured on the line profile on the direct image (blue) is about 130 nm. The same strand measured on the reflected image (red) is about 152 nm wide, which indicates slight beam broadening after reflection. More significantly, however, the reflected image shows a one-sided halo to the right side of the sample as evident from the micrograph in Fig. \ref{Fig4}(c) and the intensity level of the reflected line profile (red) to the right of each peak in Fig. \ref{Fig4}(d). This one-sided halo is resulted from a distorted scanning probe which we speculate could be due to a combination of astigmatism imparted onto the reflected beam by a slightly misaligned mirror system and the residual electric field from the mirror system near the grounded sample.

The lower contrast and brightness of the reflected image compared to the direct image, as visually apparent from micrographs as well as from the line scans shown in Fig. \ref{Fig4}(d), is largely attributed to the lower detection efficiency of secondary electrons that originated from the backside of the sample. When the in-lens secondary electron detector in our Zeiss SEM is activated, the electric field of the suction tube in the vicinity of the objective lens pole piece accelerates the secondary electrons upward towards the detector placed inside the SEM column. However, in our imaging scheme, the secondary electrons that were generated by the reflected beam on the backside of the sample have to go through a U-turn while the ones from the top surface of the sample have a more direct path to the in-lens detector. The longer, indirect path of the secondary electrons contributing to the reflected image leads to a lower number of them making it to the detector causing the lower brightness and contrast of the reflected image.

\begin{figure}[h] 
	\centering
	\includegraphics[width=1\textwidth]{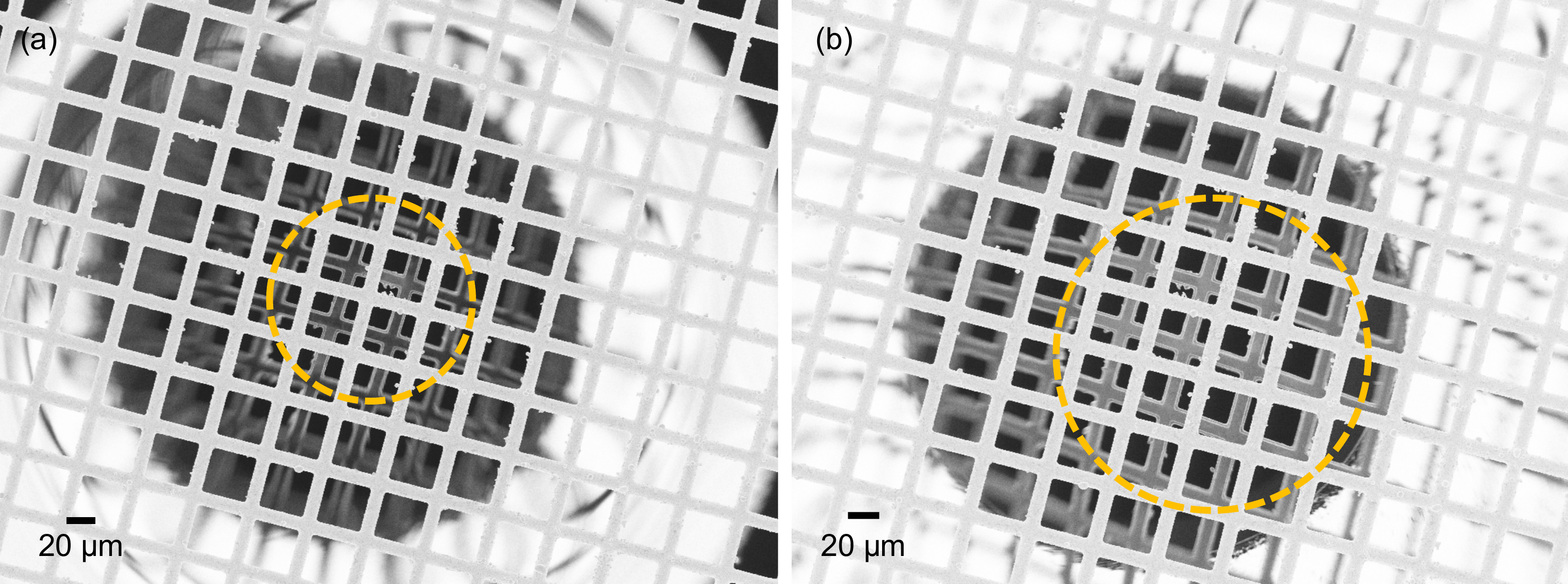}
	\caption{Approximating the area over which the reflected image remains sharp, what we define as the reflected image’s useful field of view. The brighter grid in the foreground is the direct image and the darker background grid is the reflected image of the bottom side of the sample. Imaging was conducted under the regime where the back-focal plane of the mirror system coincides with the sample plane. (a) Reflection imaging using a tetrode electron mirror setup. The yellow dashed circle points towards a useful field of view of about 120 $\upmu$m in diameter. (b) Reflection imaging using a five-electrode aperture mirror. The useful field of view as indicated by the yellow dashed circle is approximately 200 $\upmu$m in diameter.}
	\label{Fig5}
\end{figure}

The field of view in an SEM image is set by how far from the optical axis the scan coils deflect the incident probe. In our imaging scheme, the area over which the reflected beam scans the bottom side of the sample is coupled to the direct beam scan. However, downstream of the sample, where the beam enters the mirror system, the scan causes the beam to deviate from the optical axis, leading to more aberrations in the reflected probe as a function of scan angle. Consequently, the reflected image has a smaller useful field of view as shown in Fig. \ref{Fig5}(a). As indicated by the circle in Fig. \ref{Fig5}(a), the acceptable sharpness in the reflected image is limited to an area of about 120 $\upmu$m while the direct image is sharp over the entire frame, in this case 560 $\upmu$m. We observe an improvement in the useful field of view of the reflected image when using a five-electrode aperture mirror system as opposed to the flat tetrode mirror. In this case, as indicated by the dashed circle shown in Fig. \ref{Fig5}(b), the useful field of view is 200 $\upmu$m. We attribute this improvement to partially diminished aberrations of the mirror system achieved by the curved mirror potential surface.

Another practical limit to the field of view of the reflected image comes from the white halo appearing outside of the central region of the image as shown in Fig. \ref{Fig5}. What causes this bright halo is the divergent reflected beam striking the bottom of the objective lens pole piece of the SEM after passing through the image plane and generating a large number of secondary electrons which are in turn detected by the in-lens detector. This halo lowers the signal to noise ratio and places a practical limit on the field of view of the reflected image.

\begin{figure}[h] 
	\centering
	\includegraphics[width=0.6\textwidth]{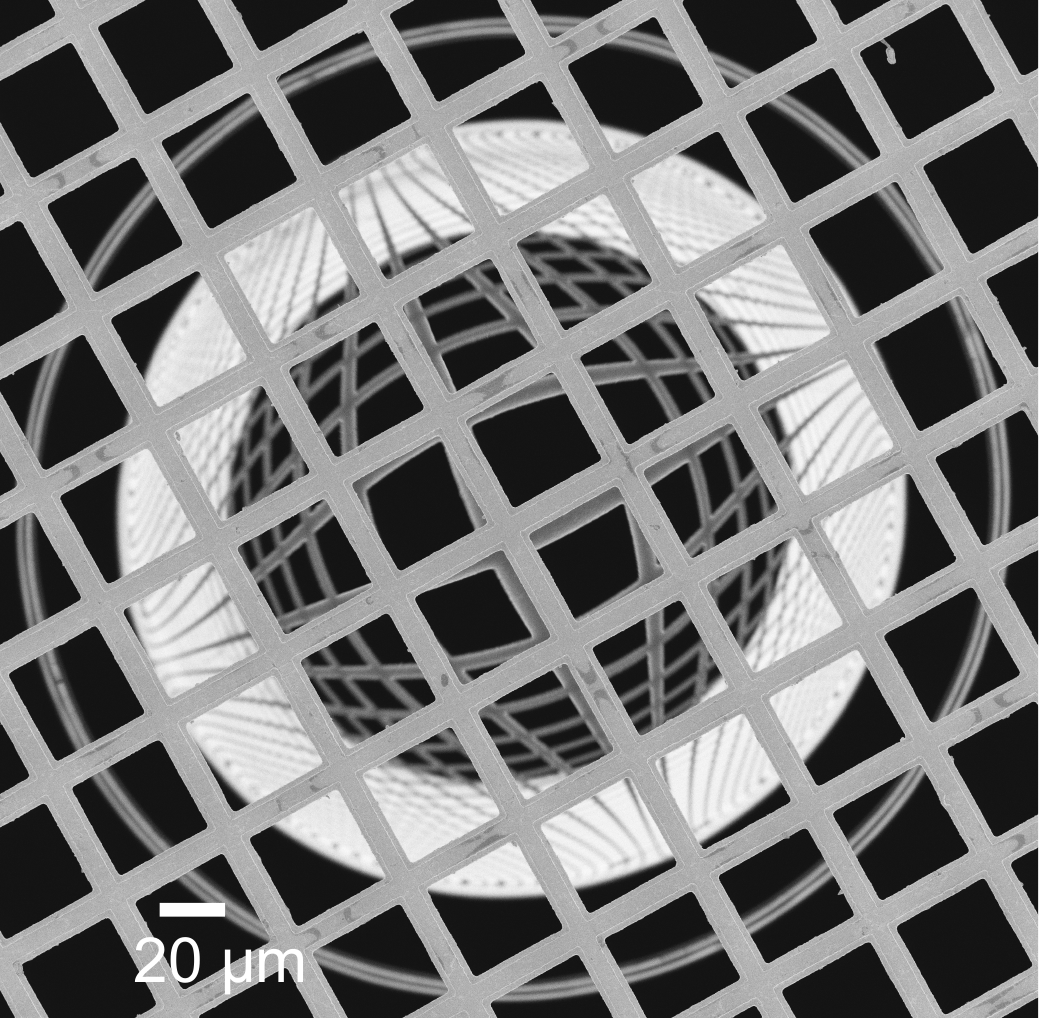}
	\caption{Demagnification of the reflected image when using a five-electrode aperture mirror under the regime where the image plane of the mirror system coincides with the sample plane. The foreground shows the direct image and the background shows the reflected image. The reflected image of features farther away from the center are demagnified more severely, leading to appearance of barrel distortion.}
	\label{Fig6}
\end{figure}

The relative magnification of the direct image compared to the reflected image is 1:1 for the first regime of operation where the sample overlaps the back-focal plane of the mirror system in both the tetrode mirror setup and the five-electrode aperture mirror setup (see Fig. \ref{Fig5}). However, as seen in Fig. \ref{Fig6}, there is considerable demagnification of the reflected image when the sample overlaps the image plane of the mirror system in the five-electrode aperture mirror setup. In this regime, a focused probe scans the surface of the concave mirror and hence there is a nonlinear relationship between the scan angle of the incident and the reflected beams. This nonlinear relationship, defined by the curvature of the reflection plane, is generated by the spherical aberration of the aperture mirror. Note that the reflected image of features farther away from the optical axis are more severely demagnified leading to the appearance of barrel distortion in the reflected image. A potential application of this effect could be probing of the curvature of the mirror for diagnostics and characterization of its spherical aberration.

\section{Conclusion and future work}
We demonstrated a scheme for simultaneous imaging of top and bottom surfaces of a sample using multi-electrode electron mirrors in an SEM. We believe that this technique could be of value to the in-situ SEM community where simultaneously observing dynamic changes on both sides of a sample may be of interest. We analyzed the image resolution currently attainable in our setup and identified ways through which this resolution could be improved. Most crucially, by imaging the pivot point of the SEM onto the reflection center, we hope to improve the resolution of the reflected image in areas farther away from the optical axis.

Next, by slightly altering the mirror setup, we demonstrated that the reflected image resolution could be improved when using a five-electrode concave mirror system. We speculate that the smaller aberrations of the curved mirror could be responsible for this improvement. We believe that for the special case of perforated samples, our imaging scheme could allow for ``in-line" spherical aberration correction in an SEM without the need for beam separators of any kind. Rigorous analysis as well as fulfillment of the aforementioned design improvements to the experimental setup are necessary for realization of this aberration correction strategy. 

In parallel, we have been developing tetrode electron mirrors fabricated using MEMS fabrication techniques. This alternative approach enables the manufacturing of very round lens apertures that eliminate the need for a stigmator to correct for the non-roundness of apertures that is encountered in conventionally machined electrode apertures \cite{Kruit2007}. Furthermore, the miniaturization of electron mirrors would allow for easier integration in a wider range of commercial SEMs. Finally, due to the smaller electrode aperture and electrode separation of these micromachined mirror systems, it would be possible to maintain similar electric fields as used in this work but with smaller applied voltages. This could open up the possibility of using accelerating lensing as opposed to decelerating lensing which would reduce the spherical aberration for a given focal length.

The potential applications of the imaging scheme reported in this work are not limited only to microscopy; indeed, it is feasible to envision this technique being extended to focused ion-beam milling. Considering that ion beams, much like electron beams, could be reflected and refocused using an electrostatic multi-electrode mirror system, one could incorporate a multi-electrode mirror under the sample in an ion-milling system. By programming the scan, the user could achieve ion milling on the top and bottom surfaces of a sample without the need to flip the sample over. Furthermore, since most focused ion beam tools come equipped with an electron column, one could envision using our mirror imaging technique to monitor the bottom surface of the specimen for signs of damage or redeposition while the sample is being milled on the top surface.

\noindent \textbf{Acknowledgments}

This work was supported by Gordon and Betty Moore Foundation and the Natural Sciences and Engineering Research Council of Canada. This research was part of the Quantum Electron Microscopy project and we would like to thank the entire QEM team (MIT, Delft University of technology, Stanford University, and University of Erlangen). We are grateful to our MIT colleagues Marco Turchetti, Akshay Agarwal, Marco Colangelo, Phillip Keathley, Brenden Butters, and Emma Batson for valuable discussions. Finally we would like to acknowledge Ben Caplins and Ryan M. White of NIST for suggesting the FIB application for our imaging scheme.

\bibliography{Mendeley_mirror_paper}
\bibliographystyle{ieeetr}

\end{document}